\documentclass[aps,prl,groupedaddress,twocolumn]{revtex4-1}
\usepackage[dvips]{graphicx}
\usepackage{dcolumn}
\usepackage{bm}
\usepackage{amsmath,amsthm,amssymb}
\usepackage{epstopdf}

\begin{document}
\title[Interaction enhanced imaging of individual atoms in dense gases]{Interaction enhanced imaging of individual atoms embedded in dense atomic gases}
\author{G. G\"unter, M. Robert-de-Saint-Vincent, H. Schempp, C. S. Hofmann}
\author{S. Whitlock}\email{whitlock@physi.uni-heidelberg.de}
\author{M. Weidem\"uller}
\affiliation{Physikalisches Institut, Universit\"at Heidelberg, Philosophenweg 12, 69120, Heidelberg, Germany.}

\newcommand{\var}{\mathrm{var} }
\newcommand{\NA}{N_{ph}^{(A)}}
\newcommand{\NB}{N_{ph}^{(B)}}
\newcommand{\NR}{N_{r}}
\newcommand{\Chi}{\mathrm{Im}[\chi]}
\newcommand{\ChiA}{\mathrm{Im}[\chi_A]}
\newcommand{\ChiB}{\mathrm{Im}[\chi_B]}
\newcommand{\ChiMax}{\mathrm{Im}[\chi_{max}]}


\begin{abstract}
We propose a new all-optical method to image individual atoms within dense atomic gases. The scheme exploits interaction induced shifts on highly polarizable excited states, which can be spatially resolved via an electromagnetically induced transparency resonance. We focus in particular on imaging strongly interacting many-body states of Rydberg atoms embedded in an ultracold gas of ground state atoms.  Using a realistic model we show that it is possible to image individual impurity atoms with enhanced sensitivity and high resolution despite photon shot noise and atomic density fluctuations. This new imaging scheme is ideally suited to equilibrium and dynamical studies of complex many-body phenomena involving strongly interacting atoms. As an example we study blockade effects and correlations in the distribution of Rydberg atoms optically excited from a dense gas.
\end{abstract}

\maketitle


The ability to prepare and probe individual quantum systems in precisely controlled environments is a driving force in modern atomic, molecular and optical physics. Manipulating single atoms~\cite{Meschede2006}, molecules~\cite{Moerner2007} and ions~\cite{Leibfried2003}, for example, is becoming a common practice. At the heart of these experiments are the powerful imaging techniques which have taken on great importance in diverse areas, such as chemical sensing and chemical reaction dynamics~\cite{Betzig1993,*Xie1994}, probing superconducting materials~\cite{Yazdani1997,*Pan2000}, and for quantum logic and quantum information processing~\cite{Schrader2004,*Nelson2007,*Haffner2005}. More recently, new single atom and single site sensitive imaging techniques for optical lattices have opened the door to control and probe complex many-body quantum systems in strongly correlated regimes~\cite{Gericke2008,*Bakr2009,*Weitenberg2011}.

The usual approach to detect atoms is to measure the fluorescence or absorption of light by driving a strong optical cycling-transition. Weak or open transitions present a difficulty since the maximum number of scattered photons per atom becomes greatly limited. In the case of long lived states of trapped ions, the technique of electron shelving has been used as an amplifying mechanism in order to directly observe quantum jumps~\cite{Nagourney86}. Another approach involves the use of an optical cavity to enhance the interaction of the atoms with the light field~\cite{Bochmann2010,*Gehr2010,*Brahms2011}. This makes it possible to reach single-atom sensitivity, but usually at the expense of greatly reduced spatial resolution.

\begin{figure}[t!]
\centering\includegraphics[width=0.88\columnwidth]{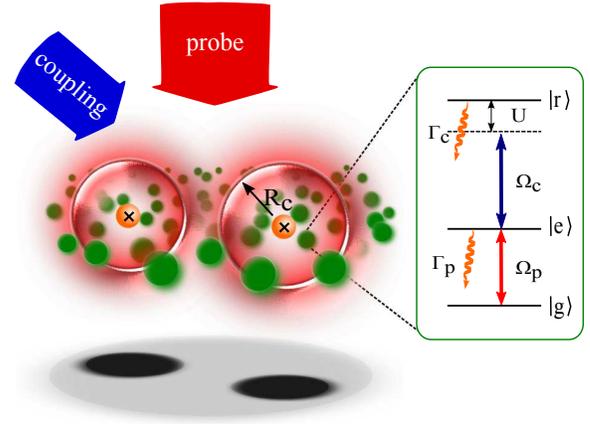}
\caption{{\bf Scheme for imaging individual impurity atoms within a dense atomic gas.} Impurity atoms (crosses) are embedded within a dense two-dimensional atomic gas of background atoms. The background atoms interact with two light fields (coupling and probe) via a two-photon resonance with an excited state $|r\rangle$. This coupling produces an EIT resonance on the ground-state probe transition. However, strong interactions with an impurity atom lead to a frequency shift $U$ of the resonance within a critical radius $R_c$. The change in absorption properties of many surrounding atoms makes it possible to map the impurity atom distribution to the absorption profile of a probe laser for analysis.} \label{fig:scheme}
\end{figure}

Here we propose a new method to image individual atoms embedded within a dense atomic gas. The concept exploits strong interactions of the atoms with highly polarizable Rydberg states of the surrounding gas. The induced level shifts can then be transferred to a strong optical transition and to many surrounding atoms within a critical radius, thereby providing two mechanisms which greatly enhance the effect of a single impurity on the light field. The Rydberg states could act as non-destructive probes for individual trapped ions, nearby surface charges, dipolar molecules, or other Rydberg atoms. In our approach, the interaction-induced shifts are spatially resolved via an electromagnetically-induced-transparency (EIT) resonance involving a weak probe and a strong coupling laser in a ladder configuration~\cite{Fleischhauer2005}. Even though the Rydberg state is barely populated, the EIT resonance is extremely sensitive to its properties~\cite{Mohapatra07,*Weatherill2008,*Pritchard2010,*Schempp2010,Tauschinsky2010}, thereby providing the means to obtain a strong absorption signal and great sensitivity combined with high spatial resolution for detecting individual atoms.

We exemplify our imaging scheme for the specific case of probing many-body states of strongly-interacting Rydberg atoms in a quasi-two-dimensional atomic gas (depicted in Fig.~\ref{fig:scheme}). Rydberg atoms are of great interest because their typical interaction ranges are comparable to, or larger than, the typical interatomic separations in trapped quantum gases. Traditionally Rydberg atoms are field ionized and the resulting ions are subsequently detected, which provides rather limited spatial resolution. As a result, much of the work done so far, such as the scaling laws for excitation~\cite{Low2009}, excitation statistics~\cite{Liebisch2005,*Amthor2010,*Viteau2011} and light-matter interactions~\cite{Mohapatra07,*Weatherill2008,*Pritchard2010,*Schempp2010}, has been restricted to the study of cloud averaged properties.  M\"uller \emph{et al.} proposed to use a single Rydberg atom to conditionally transfer an ensemble of atoms between two states~\cite{Muller2009}. Our method exploits the strong Rydberg interactions with a background gas of atoms to realize non-destructive single-shot optical images of Rydberg atoms with high resolution and enhanced sensitivity. We anticipate this technique will complement the new optical lattice imaging techniques~\cite{Gericke2008,Bakr2009,*Weitenberg2011}, but with the capability to directly image many-body systems of Rydberg atoms. We show in particular that this will provide immediate experimental access to spatial correlations in recently predicted crystalline states of highly excited Rydberg atoms~\cite{Weimer2008,*Pohl2010,*Schachenmayer2010,*vanbijnen2011}.

To quantitatively describe the absorption of probe light by a background gas of atoms surrounding a Rydberg atom we follow an approach based on the optical Bloch equations~\cite{Fleischhauer2005}. The Hamiltonian describing the atom-light coupling is
\begin{eqnarray}
\label{eq1}
H_0 &=& \frac{\hbar}{2}\big(\Omega_p | e \rangle \langle g | + \Omega_c | r \rangle \langle e |  \nonumber \\
&&+\Delta_p | e \rangle \langle e | + (\Delta_p+ \Delta_c) | r \rangle \langle r | + \textrm{h.c.}\big).
\end{eqnarray}
For resonant driving $\Delta_p=\Delta_c=0$ a dark-state is formed $|dark\rangle\approx\Omega_c|g\rangle-\Omega_p|r\rangle$, which no longer couples to the light field. Consequently, the complex susceptibility $\chi$ of the probe transition vanishes and the atoms become transparent.

The presence of a nearby Rydberg atom, however, causes an additional energy shift $U=\hbar C_6/|d|^6$ for the state $|r\rangle$, where $d$ is the distance to the Rydberg atom and the interaction coefficient $C_6$ reflects the sign and strength of interactions on the $|r\rangle$ state.
One should also account for interactions between atoms in state $|r\rangle$, but these can be neglected for $\Omega_p\ll \Omega_c$ when the population in $|r\rangle$ becomes small. We also include spontaneous decay from the states $|e\rangle$ and $|r\rangle$ with rates $\Gamma_p$, and $\Gamma_c$ respectively. From the master equation for the density matrix $\rho$ we calculate the steady-state absorption and solve for the complex susceptibility of the probe transition numerically.

In the weak probe limit ($\Omega_p\ll \Omega_c, \Gamma_{p}$) we assume the population stays mostly in the ground state ($\rho_{gg}\approx 1$). In this case we obtain for the susceptibility
\begin{eqnarray}\label{eq:chi}
\chi_=\frac{i\Gamma_p}{(\Gamma_p-2i\Delta_p)+\Omega_c^2(\Gamma_c-2i\Delta)^{-1}},
\end{eqnarray}
where $\Delta=\Delta_p+\Delta_c+C_6/|d|^6$.

\begin{figure}[t!]
\centering
\includegraphics[height=0.45\columnwidth]{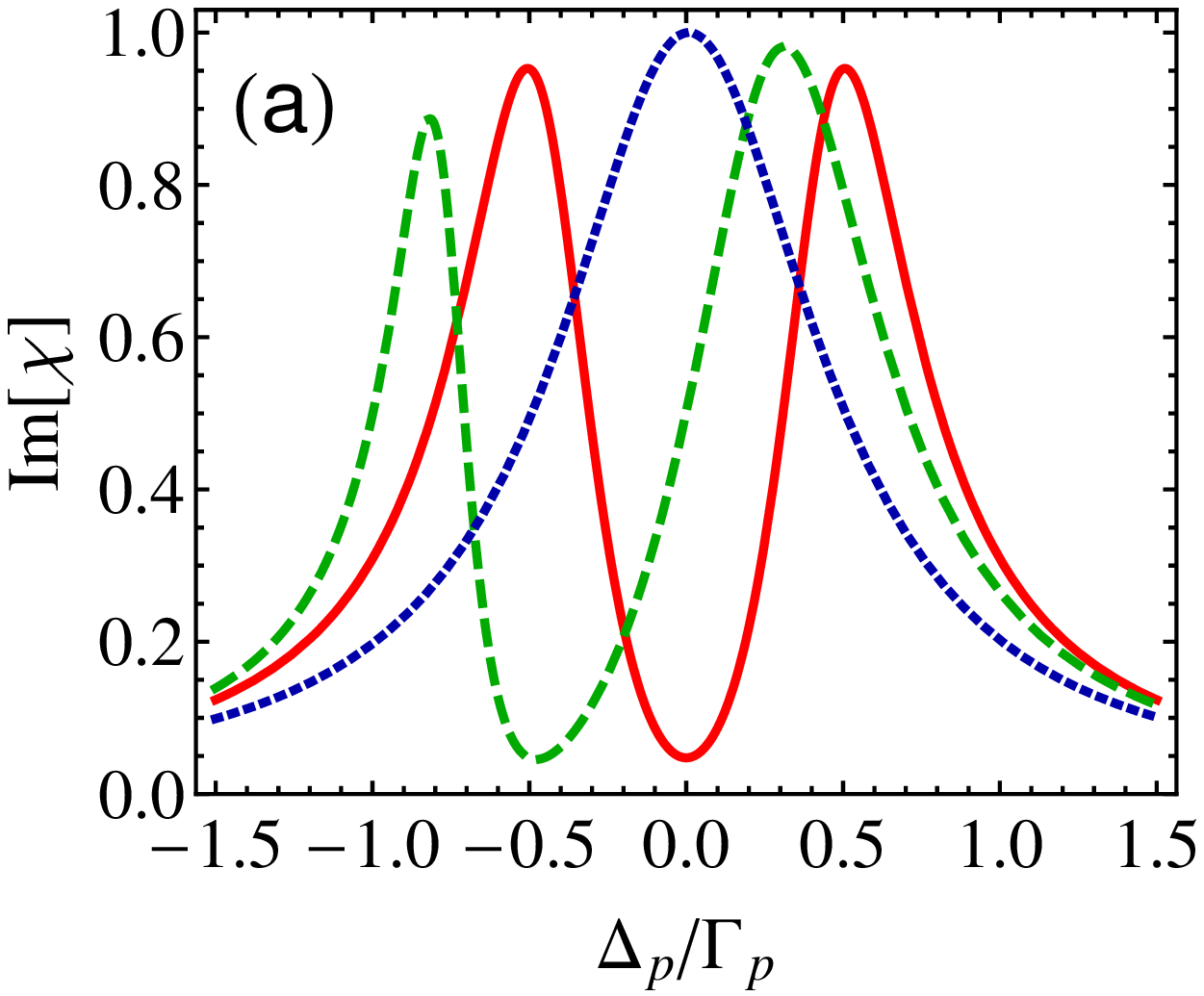}
\includegraphics[height=0.45\columnwidth]{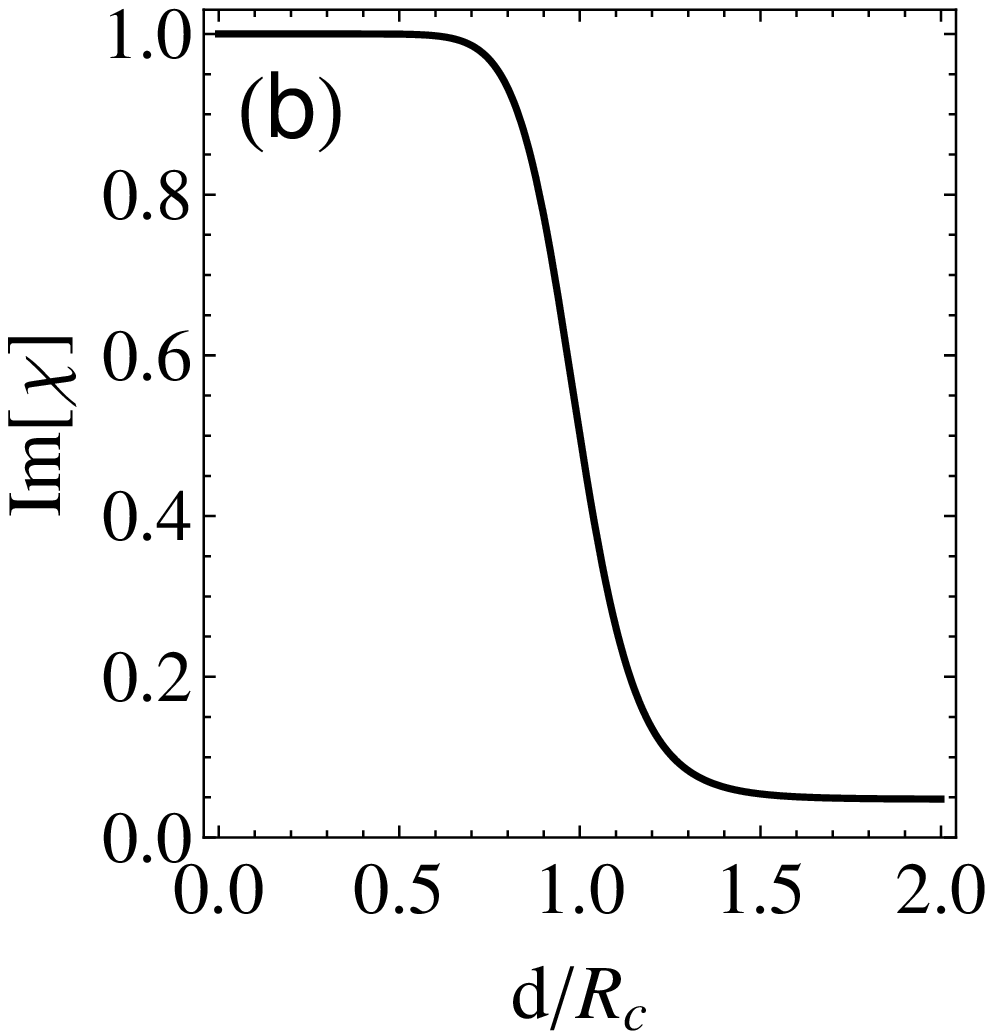}
\caption{{\bf Probe absorption given by the imaginary part of the susceptibility.} (a) $\mathrm{Im}[\chi]$ as a function of probe detuning for $\Omega_c=1$, $\Gamma_c=0.05$ and $\Delta_c=0$ (in units of $\Gamma_p$) for various distances from the Rydberg atom. The solid line is for $d\rightarrow\infty$, dashed corresponds to $d=R_c$ and the dotted line is for $d=R_c/2$. (b) Dependence of $\mathrm{Im}[\chi]$ as a function of distance from the Rydberg atom with $\Delta_p=0$. } \label{fig:susceptibility}
\end{figure}

Fig.~\ref{fig:susceptibility} shows the probe absorption proportional to the imaginary part of $\chi$ for different laser parameters and for different distances to a Rydberg atom. Far from the influence of the Rydberg atom ($d\rightarrow\infty$), the susceptibility takes on a characteristic shape with vanishing absorption on resonance. For shorter distances, interactions tend to shift the transparency window and the on-resonant susceptibility increases. At a critical distance $d=R_c$, $\Chi=1/2$. For $d<R_c$, the excited states $|r\rangle$ become far detuned and the background atoms effectively act as two-level systems. In this case the usual Lorentzian lineshape is recovered with maximum absorption on resonance. Hence, in the presence of a Rydberg atom, $N=n_{2D}\pi R_c^2\gg 1$ atoms, each with an absorption cross-section $\sim\lambda^2$, can scatter many photons to produce a dark disk with radius $R_c$. From Eq.~\ref{eq:chi} and taking $\Delta_p=0$ and $\Gamma_c\approx0$ we find the critical distance $R_c=(2 C_6\Gamma_p/\Omega_c^2)^{1/6}$, and the distance dependent susceptibility $\mathrm{Im}[\chi]\approx\big(1+(d/R_c)^{12}\big)^{-1}$. Since for typical parameters $N\approx 50$, and $R_c\approx 1~\mu$m comparable to the optical resolution, the spatially resolved probe absorption provides an excellent signature for the presence of a Rydberg atom within a dense gas.

To apply this scheme to realistic situations one also has to analyze the influence of noise. We identify two major sources, one can be attributed to photon shot noise of the probe while a second contribution is associated with intrinsic atomic density fluctuations. Both noise sources can be accurately described as Poissonian processes. This suggests that the signal-to-noise ratio can be made arbitrarily large for large intensity and large density. However, to neglect interactions between background atoms we require the probability to find more than one atom in the $|r\rangle$ state within the range of background-background interactions $R_c'$ to be $\lesssim 1$ (for the states we consider $R_c'\approx R_c$). This constraint imposes a relationship between the maximum density of background atoms and the maximum probe intensity $n_{2D} \lesssim \Omega_c^2/\pi R_c'^2\Omega_p^2$. Above this critical density the contrast of the image would decrease due to the blockade effect, where only one atom can contribute to the EIT signal, while the remainder act as two-level atoms and couple resonantly to the probe laser\cite{Pritchard2010,Ates2011}. In general, the maximum signal-to-noise ratios (see supplementary material) are achieved for large coupling strengths $\Omega_c$ and long exposure times $\tau$, but in practice these will be limited by the available laser power and by the required time resolution, which should be compared to the typical lifetime of a Rydberg atom ($\sim100~\mu$s).

\begin{figure}[t!] 
\centering
\includegraphics[width=0.49\columnwidth]{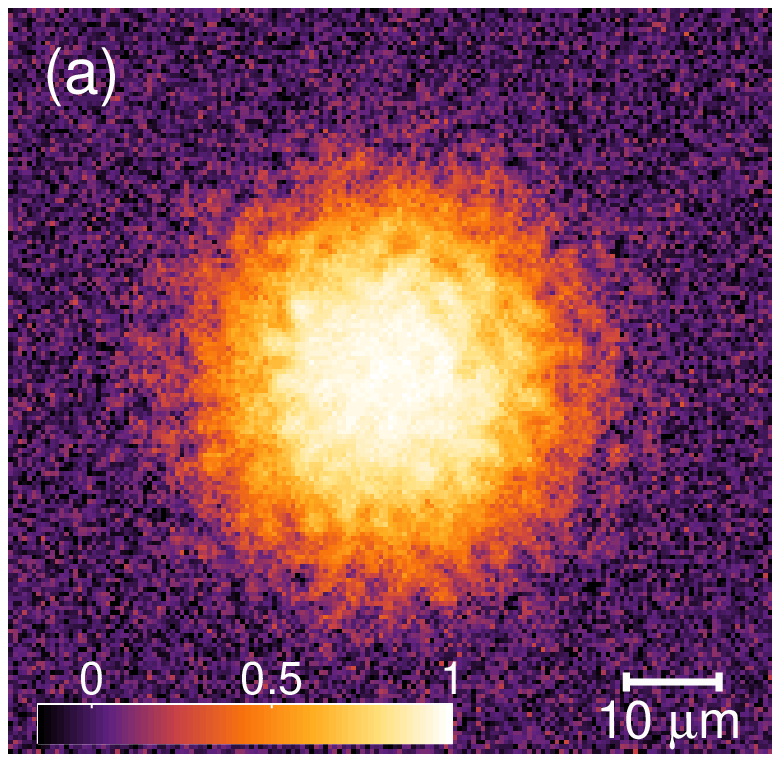} 
\includegraphics[width=0.49\columnwidth]{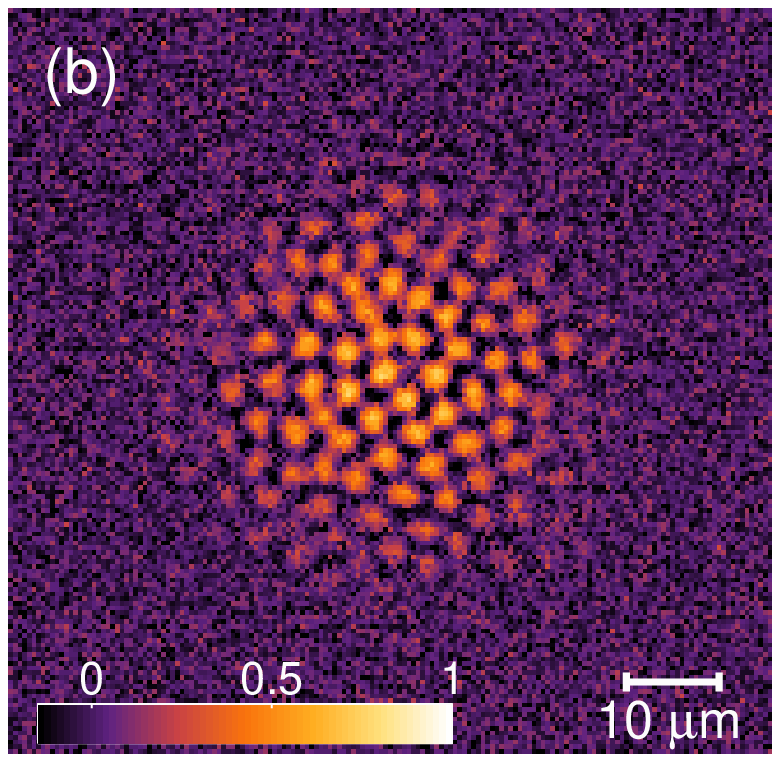} 
\caption{{\bf Simulated absorption images of atom distributions including photon shot noise and atomic density fluctuations.}
In (a), without the coupling beam, a regular absorption image of the background atoms is obtained. The color code indicates absorption. With the coupling on (b) the background atoms are rendered transparent, except for those in the vicinity of a Rydberg atom. Parameters of the simulation can be found in the text.} \label{fig:EITimage}
\end{figure}

To show the potential of this imaging scheme we have carried out numerical calculations of the EIT imaging process on simulated distributions of Rydberg atoms excited from a quasi-2D ideal gas. This situation can be realized with an optical dipole trap made using cylindrically focused Gaussian beams. Of particular interest for current experiments is the possibility to directly observe strong spatial correlations between Rydberg atoms induced by interactions in an otherwise disordered gas~\cite{Robicheaux2005, Amthor2010,Weimer2008,*Pohl2010,*Schachenmayer2010,*vanbijnen2011}.

We use a simple semi-classical model to simulate the excitation of Rydberg atoms during a chirped Rydberg excitation pulse. The model is closely related to those used to describe optical control of cold collisions~\cite{Suominen1996,Wall2007}. We consider a thermally distributed gas of 25\,000 background atoms with a peak density $n_{2D} = 40\,\rm{atoms}/\mu$m$^2$ and a cloud radius of $\sigma = 10\,\mu m$. Each atom can be in either the electronic ground state or a Rydberg state. We take the strength of the Rydberg-Rydberg interactions as $2\pi\times50~$GHz$~\mu m^6$, typical of the $55S$ state~\cite{Singer2005}. The detuning of the coupling field is swept from 0 to $+200$~MHz within $6~\mu$s with an effective Rabi frequency of $\Omega= 2\pi\times2.0~$MHz. During each time step atoms can undergo a transition to the Rydberg state with a probability estimated from the Landau-Zener formula~\cite{Wittig2005}. Successive excitation events are treated independently, however the previously excited atoms influence successive transitions through the Rydberg-Rydberg interactions. This model also includes the effects of mechanical forces between Rydberg atoms by simultaneously solving the classical equations of motion for each Rydberg excited atom.

 We then calculate absorption images of the simulated Rydberg distributions by numerically solving the optical Bloch equations for the background atoms at each spatial position, accounting for the level-shifts produced by all Rydberg atoms. From the mean intensity at each pixel we generate Poisson distributed photon-shot noise. Similarly, a reference image is generated with uncorrelated noise for background division. Fig.~\ref{fig:EITimage} shows calculated absorption images without and with the coupling laser from a single run of the simulation. Each pixel corresponds to a region of $(0.5\,\mu\rm{m})^2$ in the plane of the atoms and we assume a numerical aperture of $0.25$ and an exposure time of $10\,\mu$s. For the EIT ladder system, we take the $^{87}$Rb states, $|5S_{1/2}, F=2, m_F = 2\rangle $ for the ground state, $|5P_{3/2}, F=3, m_F = 3\rangle$ for the intermediate state, and $|r = 28S\rangle$ for the excited state. The decay rates are $\Gamma_p=2\pi\times 6.1$ MHz and $\Gamma_c \approx 2\pi\times 10\,$kHz, and for the coupling laser we assume $\Omega_c = 2 \pi\times 50\,$MHz. Laser line widths of $2\pi \times 1\,$MHz were assumed for both probe and coupling lasers. The interactions between $|55S\rangle$ and $|28S\rangle$ states were calculated to obtain a van der Waals coefficient of $C_6(28S-55S) = -2\pi\times 8.7\,\rm{MHz}\,\mu \rm{m}^6$ giving $R_c = 0.59\,\mu$m. Interactions between background atoms are taken as $C_6(28S-28S) = 2\pi\times 10.1\,\rm{MHz}\,\mu \rm{m}^6$ ($R_c'=0.61\mu$m).  For these parameters the optimal signal-to-noise ratio is obtained for $n_{2D}\approx 40\,$atoms/$\mu$m$^2$. The probe intensity ($\Omega_p =2\pi \times 5.8\,$MHz) is chosen such that on average the $|r\rangle$ state density remains below 1 per $\pi R_c'^2$. Such parameters are readily achieved in current experiments with quasi-2D atomic gases~\cite{Gorlitz01}.

In the background region of the image the signal is dominated by photon-shot noise, while at the center atom-shot noise dominates. With the coupling laser on the ground state atoms are rendered mostly transparent, except for regions of high absorption around each Rydberg atom (Fig.~\ref{fig:EITimage}b). The locations of the individual Rydberg atoms are clearly resolved in the image as bright (absorbing) spots with a spatial extent of $2.3~\mu$m FWHM comparable to the assumed optical resolution. One can easily envisage higher resolutions using state-of-the-art imaging systems~\cite{Bakr2009,*Weitenberg2011}, with the fundamental limit given by the density of background atoms surrounding the impurities. The signal-to-noise ratio of our images is sufficiently high that we can fit the position of each Rydberg atom with subpixel precision.

\begin{figure}[t!]
\centering
\includegraphics[width=0.8\columnwidth]{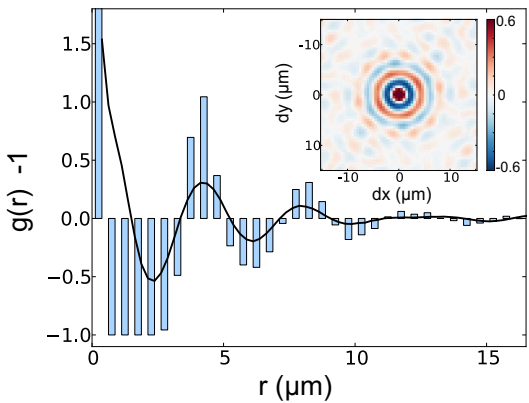}
\caption{\label{fig:correlation}
{\bf Pair distribution function computed for simulated Rydberg images from 15 realisations.} The inset shows the averaged two-dimensional autocorrelation function as computed from the absorption images. Taking the radial average gives $g(r)-1$ as shown in the main figure (solid black line). The shaded bars show $g(r)-1$ with $0.5~\mu$m bin size as obtained from the simulated Rydberg atom coordinates. The clear shell structure which reflects translational order between nearest and next-nearest neighbours is preserved by the images.
}
\end{figure}

 Despite the simplicity of the excitation model, the final distribution of Rydberg atoms appears highly-correlated, reproducing some of the features of a full quantum mechanical treatment~\cite{Pohl2010}. To characterize the translational order of the simulated Rydberg distributions, we calculate the pair distribution function $g(r)$ from 15 simulated images (Fig.~\ref{fig:correlation}). To account for the inhomogeneous density distribution we also normalize by the autocorrelation of the mean image (see supplementary material). For a random distribution of atoms $g(r) \approx 1$. Larger correlation values indicate an enhanced probability to find two Rydberg atoms at a given separation, while lower values indicate the absence of pairs. Since there is no preferred orientation in our system $g(r)$ takes on cylindrical symmetry. We clearly observe a shell with $g(r)\approx 0$ at a radius of $\sim$2.5~$\mu$m which reflects the strong blockade of excitation due to Rydberg-Rydberg interactions. At larger distances, we observe two positive-correlated shells (around 4 and 8~$\mu$m), which indicate translational correlations between nearest- and next-nearest neighbours. The observed shell structure decays rapidly indicating the absence of true long-range order. We note very similar behaviour of $g(r)$ for the raw atom positions (shaded bars). From this we conclude that the information regarding density-density correlations can be reliably extracted from the images, even under realistic imaging conditions.

We have also computed the angular correlation function $\Phi(\theta)$ at the radius of the first shell (see supplementary material). This gives the probability, starting from an atom to find two neighbours forming an angle $\theta$. We observe the presence of two peaks at $\sim$1.1 and $\sim$5.3 rad, corresponding closely to $\pi/ 3$ and $5 \pi / 3$, reflecting the 6-fold symmetry present among most nearest-neighbours. Even more information could be obtained from these images by studying higher-order correlation functions, or by first estimating the Rydberg atom positions to fully characterize the many-body state.

Our new imaging method provides the means to optically image individual atoms within a dense atomic gas using Rydberg state electromagnetically induced transparency. The conditions we find to optimise the signal closely match those of current cold atom experiments. As an example of the full potential of this new imaging scheme we have carried out numerical simulations of Rydberg atoms excited from a quasi-2D gas. Remarkably, this simple model already gives rise to the appearance of strong spatial correlations between Rydberg atoms. The EIT imaging scheme provides the means for single-shot, non-destructive and time resolved images of such many-body states. We can foresee numerous other applications of the EIT imaging method. For example, Rydberg state EIT has already been used to measure spatially inhomogeneous electric fields near a surface~\cite{Tauschinsky2010}. Another exciting prospect would be to directly image single ions within an atomic gas which could be directly observed in current experiments~\cite{Grier2009,*Zipkes2010,*Schmid2010}. Closely related ideas could be used to realise a single atom optical transistor~\cite{Hwang2009} or to impose non-classical spatial correlations onto the light field~\cite{Sevlinci2011}.

\acknowledgements{ We would like to thank J. Evers, B. Olmos and I. Lesanovsky for valuable discussions. This work is supported in part by the Heidelberg Center for Quantum Dynamics and the Deutsche Forschungsgemeinschaft under WE2661/10.1. SW acknowledges support from the EU Marie-Curie program (grant number PERG08-GA-2010-277017).}

\emph{Note added.--} During preparation of this manuscript we became aware of related work~\cite{Olmos2011}

\section{Supplementary material}
\subsection{Influence of noise}

We provide the criteria to optimize the quality of images in the presence of realistic noise sources. Suppose an impurity atom is situated in region $(A)$ and we wish to distinguish its position from another region $(B)$ by measuring the difference in transmission $ \Delta T$. $N_{ph}^{(A/B)}$ is the number of detected photons (proportional to CCD counts; we assume a quantum efficiency $\approx1$) and $\NR$ is a reference used to normalize probe intensity variations. We consider the resonant case ($\Delta_p=\Delta_c=0$) since this maximizes the contrast while keeping the density of atoms in state $|r\rangle$ small. In region $(A)$ the absorption is greatly enhanced and $N_{ph}^{A}\approx T_A\NR$ with $T_A=\exp{(-\sigma_0 n_{2D} \Chi)}$ and $\sigma_0$ is the resonant absorption cross-section for the probe transition. In region $(B)$ $\ChiB\approx 0$ therefore $\Delta T >0 $.

The precision with which a measurement of $\Delta T$ can be made depends on the noise in both regions:
\begin{eqnarray}\label{eq:varT1}
\var(\Delta T)\!\approx\! \frac{\var(\NR)\langle\NA\rangle^2}{\langle\NR\rangle^4}\!+\!\frac{\var(\NA)}{\langle\NR\rangle^2}\!+\!\frac{2 \, \var(\NR) }{\langle\NR\rangle^2}.\nonumber
\end{eqnarray}
We assume Poisson distributed noise for the intensity and density fluctuations, so $\var(\NR)=\nobreak\langle\NR\rangle$ and $\var(N_{ph})\approx\nobreak\langle T_A\rangle\langle\NR\rangle+\langle\NR\rangle^2\var(T_A)$. Atom shot noise is accounted for by $\var(T_A)=\nobreak\sigma_0^2 \ChiA^2\langle T_A\rangle^2 n_{2D}/ a$, with $a$ the area of each region (for example the area of a pixel),
\begin{eqnarray}
\var(\Delta T)\!=\!\frac{\langle T_A\rangle\!+\!\langle T_A\rangle^2}{\langle\NR\rangle}\!+\!  \frac{2}{\langle\NR\rangle}\!+\!\frac{\sigma_0^2n_{2D}}{a} \ChiA^2\langle T_A\rangle^2.\nonumber
\end{eqnarray}
The first two terms can be attributed to photon shot noise while the last term is from density fluctuations. Including saturation, $\ChiA=\nobreak\Gamma_p^2/(\Gamma_p^2+\nobreak 2 \Omega_p^2)$. This suggests that the signal-to-noise ratio (SNR) can be made arbitrarily high for large $\langle \NR\rangle$ and large $n_{2D}$. However, to ensure that interactions between background atoms can be neglected, we require that the density of atoms in the $|r\rangle$ state is kept low ($\rho_{rr}n_{2D}\pi R_c'^2\lesssim 1$). For strong coupling $\rho_{rr}\approx \Omega_p^2/\Omega_c^2$ and this implies $\langle\NR\rangle\lesssim\nobreak a \tau \Omega_c^2/\sigma_0n_{2D} \pi R_c'^2\Gamma_p$, with exposure time $\tau$. In the limit of strong absorption $\langle T_A\rangle\ll 1$, and substituting for the maximum value of $\langle\NR\rangle$:
\begin{eqnarray}\label{eq:varfinal}
\var(\Delta T)&=& \frac{2 \sigma_0 \Gamma_p n_{2D}\pi R_c'^2}{a \Omega_c^2 \tau}
\\ \nonumber
&\times &
\biggr(1+\frac{\Omega_c^2 \tau \sigma_0}{2 \pi \Gamma_p R_c'^2}\ChiA^2\exp{(-2 \sigma_0 n_{2D} \ChiA})\biggr)
\end{eqnarray}
with $\ChiA=\big(1+2\Omega_c^2/\Gamma_p^2 \pi R_c'^2 n_{2D}  \big)^{-1}$.

In general, the best SNR is obtained for large coupling strengths $\Omega_c$ and long exposure times $\tau$, but in practice these will be limited by the available laser power and by the required time resolution. To find the optimal values for $n_{2D}$ and $\Omega_p$ given fixed values of $\tau$ and $\Omega_c$ we numerically maximize the SNR using Eq.~\eqref{eq:varfinal}. The final parameters used in the paper include the additional effect of finite laser linewidths which tends to increase $\rho_{rr}$ slightly for the same $\Omega_p$. This shifts the optimum density to slightly lower values. Assuming $\Omega_c=2\pi\times 50$~MHz and $\tau=10~\mu$s we find $n_{2D}^{opt}=40 \mu$m$^{-2}$ (neglecting linewidth $n_{2D}^{opt}\approx 50 \mu$m$^{-2}$).

\subsection{Rydberg excitation model}

To simulate the excitation of Rydberg atoms by a chirped laser pulse we consider a randomly (thermally) distributed ensemble of atoms. Each atom is treated as a point-like classical particle which can be in either the electronic ground state or in a Rydberg state. As the coupling field is swept from low to high detuning, each atom can undergo a transition. The transition probability is estimated using the Landau-Zener (LZ) formula for a sweep through an avoided crossing \cite{Wittig2005}. The effect of Rydberg-Rydberg interactions causes level shifts for the nearby atoms which subsequently alters their probability to be excited by the laser pulse, giving rise to strong spatial correlations.

The simulation starts with zero detuning for the excitation laser and one atom is chosen at random to start in the Rydberg state. In the next time step the laser frequency is varied according to a fixed sweep rate, and we calculate all level shifts due to Rydberg-Rydberg interactions. From the atoms which crossed the resonance condition in the previous timestep we randomly select newly excited atoms based on their LZ probabilities. Any successful excitation immediately influences all other surrounding atoms, and thus the simulation also reproduces the excitation blockade effect. For each time step we also solve the Newtonian equations of motion of the Rydberg atoms to account for the interparticle mechanical forces. We do not consider the motion of the ground state atoms for the simulation (frozen gas regime). The simulation returns a list of the final coordinates of all the ground-state and Rydberg atoms within the gas after the laser sweep. These coordinates are then used as inputs to calculate the corresponding absorption image.

\subsection{Correlation analysis}

To characterize the translational order of the simulated Rydberg distributions, we define a pair distribution function from the absorption images $n(\vec r)$:
\begin{equation}
G[n](\vec r) = \frac{ \int d^2r_0  ~n(\vec r_0) n(\vec r_0 + \vec r)} { \left( \int d^2r_0 ~ n(\vec r_0)    \right) ^2}.
\end{equation}
To account for the inhomogeneous density and finite size of the system we define the following rescaled pair distribution function :
\begin{equation}
g(\vec r)  = \frac{\langle G_2[n] \rangle }{ G_2[\langle n\rangle] }
\label{eq:g2}
\end{equation}
where the brackets reflect averages over independent realisations. For a random distribution of atoms $g(r) \approx 1$. Larger correlation values indicate an enhanced probability to find two Rydberg atoms at a given separation, while lower values indicate the absence of pairs.

\begin{figure}[t!]
\centering
\includegraphics[width=0.8\columnwidth]{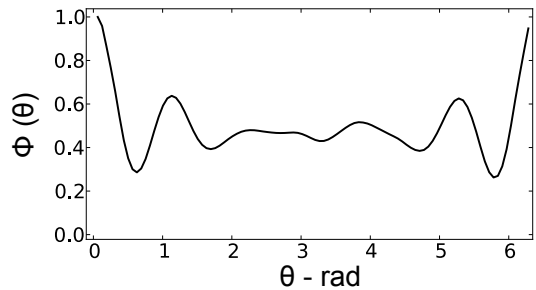}
\caption{\label{fig:correlation2}
{\bf Angular correlation function computed from 15 simulated images.} Angular correlation function $\Phi(\theta )$ taken at the radius of the first shell. We observe peaks at angles around $\pm \pi / 3$ indicating a six-fold symmetry among nearest neighbours.
}
\end{figure}

We can also extract information about the angular correlations in the images. For this we define the angular correlation function :
\begin{equation}
\Phi(\theta)\!\propto\!\left\langle\! \frac{        \int\! d^2r_0~n(\vec r_0)  \int\! d \phi~n(\vec r_0\!+\! R_{nn}\vec{e}_{\phi} ) n(\vec r_0\!+\!R_{nn} \vec{e}_{\phi+\theta} )                       }                  { \left(\int d^2r_0  ~n(\vec r_0) \right) ^3}  \!\right\rangle
\label{eq:g3}
\end{equation}
where $\vec e_{\phi}$ is defined as the unit vector with angle $\phi$ with respect to a reference axis $\vec e_x$, and $R_{nn}$ is the radius of the first positive shell of the pair distribution function. This gives the probability, starting from an atom and one of its nearest neighbours, to find a second nearest neighbour forming an angle $\theta$ with the first.

Figure~\ref{fig:correlation2} shows the angular correlation function computed from 15 simulated images at the radius of the first shell. We observe two clear peaks at $\sim \pi/3$ and $\sim 5\pi/3$ reflecting the 6-fold symmetry present among nearest neighbours. The other peaks at $\theta=2n\pi/6,n=2,3,4$ are washed out indicating the absence of true long range orientational order.

\bibliography{Gunter11}

\begin{thebibliography}{10}
\bibitem{Meschede2006}D. Meschede and A. Rauschenbeutel, Adv. At. Mol. Opt.
  Phys. {\bf 53},  75   (2006).
\bibitem{Moerner2007}W.~E. Moerner, Proc. Natl. Acad. Sci. (USA) {\bf 104},
  12596  (2007).
\bibitem{Leibfried2003}D. Leibfried, R. Blatt, C. Monroe, and D. Wineland, Rev.
  Mod. Phys. {\bf 75},  281  (2003).
\bibitem{Betzig1993}E. Betzig and R.~J. Chichester, Science {\bf 262},  1422
  (1993).
\bibitem{Xie1994}X.~S. Xie and R.~C. Dunn, Science {\bf 265},  361  (1994).
\bibitem{Yazdani1997}A. Yazdani {\it et~al.}, Science {\bf 275},  1767  (1997).
\bibitem{Pan2000}S.~H. Pan {\it et~al.}, Nature {\bf 403},  746  (2000).
\bibitem{Schrader2004}D. Schrader {\it et~al.}, Phys. Rev. Lett. {\bf 93},
  150501  (2004).
\bibitem{Nelson2007}K.~D. Nelson, X. Li, and D.~S. Weiss, Nature Phys. {\bf 3},
   556  (2007).
\bibitem{Haffner2005}H. H{\"a}ffner {\it et~al.}, Nature {\bf 438},  643
  (2005).
\bibitem{Gericke2008}T. Gericke {\it et~al.}, Nature Phys. {\bf 4},  949
  (2008).
\bibitem{Bakr2009}W.~S. Bakr {\it et~al.}, Nature {\bf 462},  74  (2009).
\bibitem{Weitenberg2011}C. Weitenberg {\it et~al.}, Nature {\bf 471},  319
  (2011).
\bibitem{Nagourney86}W. Nagourney, J. Sandberg, and H. Dehmelt, Phys. Rev.
  Lett. {\bf 56},  2797  (1986).
\bibitem{Bochmann2010}J. Bochmann {\it et~al.}, Phys. Rev. Lett. {\bf 104},
  203601  (2010).
\bibitem{Gehr2010}R. Gehr {\it et~al.}, Phys. Rev. Lett. {\bf 104},  203602
  (2010).
\bibitem{Brahms2011}N. Brahms {\it et~al.}, Nature Phys. (adv. online)  (2011).
\bibitem{Fleischhauer2005}M. Fleischhauer, A. Imamoglu, and J.~P. Marangos,
  Rev. Mod. Phys. {\bf 77},  633  (2005).
\bibitem{Mohapatra07}A.~K. Mohapatra, T.~R. Jackson, and C.~S. Adams, Phys.
  Rev. Lett. {\bf 98},  113003  (2007).
\bibitem{Weatherill2008}K.~J. Weatherill {\it et~al.}, J. Phys. B: At. Mol.
  Opt. Phys. {\bf 41},  201002  (2008).
\bibitem{Pritchard2010}J.~D. Pritchard {\it et~al.}, Phys. Rev. Lett. {\bf
  105},  193603  (2010).
\bibitem{Schempp2010}H. Schempp {\it et~al.}, Phys. Rev. Lett. {\bf 104},
  173602  (2010).
\bibitem{Tauschinsky2010}A. Tauschinsky {\it et~al.}, Phys. Rev. A {\bf 81},
  063411  (2010).
\bibitem{Low2009}R. L\"ow {\it et~al.}, Phys. Rev. A {\bf 80},  033422  (2009).
\bibitem{Liebisch2005}T.~C. Liebisch, A. Reinhard, P.~R. Berman, and G.
  Raithel, Phys. Rev. Lett. {\bf 95},  253002  (2005).
\bibitem{Amthor2010}T. Amthor, C. Giese, C.~S. Hofmann, and M. Weidem\"uller,
  Phys. Rev. Lett. {\bf 104},  013001  (2010).
\bibitem{Viteau2011}M. Viteau {\it et~al.}, arXiv:1103.4232  (2011).
\bibitem{Muller2009}M. M\"uller {\it et~al.}, Phys. Rev. Lett. {\bf 102},
  170502  (2009).
\bibitem{Weimer2008}H. Weimer, R. L{\"o}w, T. Pfau, and H. B{\"u}chler, Phys.
  Rev. Lett. {\bf 101},  250601  (2008).
\bibitem{Pohl2010}T. Pohl, E. Demler, and M.~D. Lukin, Phys. Rev. Lett. {\bf
  104},  043002  (2010).
\bibitem{Schachenmayer2010}J. Schachenmayer, I. Lesanovsky, A. Micheli, and
  A.~J. Daley, New J. Phys. {\bf 12},  103044  (2010).
\bibitem{vanbijnen2011}R.~M.~W. {van Bijnen} {\it et~al.}, arXiv:1103.2096
  (2011).
\bibitem{Ates2011}C. Ates, S. Sevin{\c{c}}li, and T. Pohl, Phys. Rev. A {\bf
  83},  041802(R)  (2011).
\bibitem{Robicheaux2005}F. Robicheaux and J. Hern{\'a}ndez, Phys. Rev. A {\bf
  72},  063403  (2005).
\bibitem{Suominen1996}K.-A. Suominen, J. Phys. B: At. Mol. Opt. Phys. {\bf 29},
   5981  (1996).
\bibitem{Wall2007}M.~L. Wall, F. Robicheaux, and R.~R. Jones, J. Phys. B: At.
  Mol. Opt. Phys. {\bf 40},  3693  (2007).
\bibitem{Singer2005}K. Singer, J. Stanojevic, M. Weidem{\"u}ller, and R.
  C{\^o}t{\'e}, J. Phys. B: At. Mol. Opt. Phys. {\bf 38},  S295  (2005).
\bibitem{Wittig2005}C. Wittig, J. Phys. Chem. B {\bf 109},  8428  (2005).
\bibitem{Gorlitz01}A. G\"orlitz {\it et~al.}, Phys. Rev. Lett. {\bf 87},
  130402  (2001).
\bibitem{Grier2009}A. Grier, M. Cetina, F. Oru{\v{c}}evi{\'c}, and V.
  Vuleti{\'c}, Phys. Rev. Lett. {\bf 102},  223201  (2009).
\bibitem{Zipkes2010}C. Zipkes, S. Palzer, C. Sias, and M. Kohl, Nature {\bf
  464},  388  (2010).
\bibitem{Schmid2010}S. Schmid, A. H{\"a}rter, and J. Denschlag, Phys. Rev.
  Lett. {\bf 105},  133202  (2010).
\bibitem{Hwang2009}J. Hwang {\it et~al.}, Nature {\bf 460},  76  (2009).
\bibitem{Sevlinci2011}S. Sevin{\c c}li, N. Henkel, C. Ates, and T. Pohl,
  arXiv:1106.2001  (2011).
\bibitem{Olmos2011}B. Olmos, W. Li, S. Hofferberth, and I. Lesanovsky,
  arXiv:1106.4444  (2011).

\end{thebibliography}

\end{document}